\documentclass[]{emulateapj}

\usepackage{graphicx}
\usepackage{epsfig}
\usepackage{txfonts}
\usepackage{amssymb}
\usepackage{epsf}
\usepackage{natbib}

\shorttitle{The Structure \& Dynamics of Massive Early-type Galaxies}
\shortauthors{Koopmans et al.}

\submitted{Accepted for publication in ApJL}

\begin{document}

\title{The Structure \& Dynamics of Massive Early-type Galaxies: \\ On Homology, Isothermality and Isotropy
	inside one effective radius}

\author{L.V.E. Koopmans\altaffilmark{1},  A. Bolton\altaffilmark{2}, T. Treu\altaffilmark{3}, O. Czoske\altaffilmark{1}, M. W. Auger \altaffilmark{3}, M. Barnab\`e\altaffilmark{1}, S. Vegetti\altaffilmark{1}, R. Gavazzi\altaffilmark{4}, L. A. Moustakas\altaffilmark{5}, S. Burles\altaffilmark{6}}

\altaffiltext{1}{Kapteyn Astronomical Institute, University of Groningen, P.O.Box 800, 9700 AV Groningen, The Netherlands ({\tt koopmans@astro.rug.nl})}
\altaffiltext{2}{Institute for Astronomy,
University of HawaiÔi, 2680 Woodlawn Dr., Honolulu, HI 96822
({\tt bolton@ifa.hawaii.edu})}
\altaffiltext{3}{Department of Physics, University of California, Santa Barbara, CA 93106, USA ({\tt ttreu@physics.ucsb.edu})}
\altaffiltext{4}{Institut dÕAstrophysique de Paris, UMR7095 CNRS - Universit\'e Paris 6, 98bis Bd Arago, 75014 Paris, France ({\tt gavazzi@iap.fr})}
\altaffiltext{5}{JPL/Caltech, MS 169-327, 4800 Oak Grove Dr., Pasadena, CA 91109 ({\tt leonidas@jpl.nasa.gov})}
\altaffiltext{6}{Department of Physics and Kavli Institute for Astrophysics and
Space Research, Massachusetts Institute of Technology, 77 Massachusetts
Avenue, Cambridge, MA 02139, USA ({\tt burles@mit.edu})}

\begin{abstract}

Based on 58 SLACS strong-lens early-type galaxies with direct total-mass and stellar-velocity dispersion measurements, we find that inside one effective radius massive elliptical galaxies with $M_{\rm eff} \ga 3\cdot 10^{10}$\,M$_{\odot}$ are well-approximated by a power-law ellipsoid with an average logaritmic density slope of $\langle \gamma'_{\rm LD} \rangle \equiv -d\log(\rho_{\rm tot})/d\log(r)=2.085^{+0.025}_{-0.018}$ (random error on mean) for isotropic orbits with $\beta_r=0$, $\pm 0.1$ (syst.) and $\sigma_{\gamma'}\la 0.20^{+0.04}_{-0.02}$ intrinsic scatter (all errors indicate the 68\% CL). We find no correlation of $\gamma'_{\rm LD}$ with galaxy mass ($M_{\rm eff}$), rescaled radius (i.e. $R_{\rm einst}/R_{\rm eff}$) or redshift, despite intrinsic differences in density-slope between galaxies. Based on scaling relations, the average logarithmic density slope can be derived in an alternative manner, {\sl fully independent from dynamics}, yielding  $\langle \gamma'_{\rm SR} \rangle =1.959\pm 0.077$. Agreement between the two values is reached for $\langle \beta_r \rangle =0.45 \pm 0.25$, consistent with mild radial anisotropy. This agreement supports the robustness of our results,
despite the increase in mass-to-light ratio with total galaxy mass: $M_{\rm eff}\propto L_{V,\rm eff}^{1.363 \pm 0.056}$.  We conclude that massive early-type galaxies are structurally close-to homologous with close-to isothermal total density profiles ($\la$10\% intrinsic scatter) and have at most some mild radial anisotropy. Our results provide new observational limits on galaxy formation and evolution scenarios, covering four Gyr look-back time. 
\end{abstract}

\keywords{gravitational lensing --- galaxies: structure}

\section{Introduction}

Understanding the internal structure of massive early-type galaxies (ETG) is essential, if we ever hope to fully understand hierarchical galaxy formation \citep[e.g.][]{1985ApJ...292..371D, 1985Natur.317..595F, 1991ApJ...379...52W} and the complex interplay between dark-matter and baryons.  A number of tight relations, such as the Fundamental Plane \citep[FP hereafter;][]{1987ApJ...313...42D, 1987ApJ...313...59D}, the color-magnitude relation \citep[e.g.][]{1977ApJ...216..214V, 1978ApJ...223..707S, 1992MNRAS.254..601B} and the relation between black-hole and stellar spheroid masses \citep{1995ARA&A..33..581K, 1998AJ....115.2285M, 2000ApJ...539L...9F, 2000ApJ...539L..13G}, indicate that there must be physical processes that are not dominated by stochastic processes of hierarchical galaxy formation. 

In the FP relation of ETGs both stars and dark-matter contribute to the
structure and dynamics ($\sigma_{\rm c}$) and total mass-to-light ratio ($M/L$) . Hence, without  knowledge of the dark-matter distribution and stellar populations, we remain clueless about whether changes in the FP are due to non-homology or due to changes in their stellar mass-to-light, or both \citep[e.g.][]{1995ApJ...445...55H, 1997A&A...321..111P, 1999MNRAS.308.1037T, 2001AJ....121.1936G, 2002A&A...386..149B, 2006MNRAS.366.1126C, 2006MNRAS.370..681N,
2009arXiv0903.3603G}. Given only first and second moments of the stellar velocity distribution, disentangling these effects has remained difficult, because of the mass-anisotropy degeneracy: steepening of the density profile and changes in the orbital anisotropy can offset each other to yield similar kinematic profiles \citep[e.g.][]{1993MNRAS.265..213G, 1998MNRAS.295..197G, 2003MNRAS.343..401L}. In studying the tilt of the FP \citep[e.g.][]{1993ApJ...416L..49R}, one can therefore not easily disentangle effects of mass structure (e.g. non-homology) from changes in stellar mass-to-light ratios, using only stellar kinematics. 

In recent papers \citep{2006ApJ...638..703B, 2006ApJ...650.1219T, 2006ApJ...649..599K, 2007ApJ...665L.105B, 2007ApJ...667..176G, 2008ApJ...682..964B}, we analyzed a subsample of well-selected gravitational lens systems from the SLACS Survey\footnote{\tt www.slacs.org}, showing that massive elliptical galaxies  have {\sl on average} close-to isothermal density profiles, with some minor, but noticeable, intrinsic scatter between their logarithmic density slopes \citep[e.g.][]{2006ApJ...649..599K, 2004ApJ...611..739T}. They follow the classical FP \citep{2006ApJ...650.1219T}, as well as a tight Mass Fundamental Plane \citep[MFP;][]{2007ApJ...665L.105B, 2008ApJ...684..248B}, where galaxy surface brightness is replaced by surface density. In all observable respects, they follow the trends of normal elliptical galaxies \citep{2006ApJ...638..703B, 2006ApJ...650.1219T, 2008ApJ...682..964B, 2009ApJ...690..670T} and these lens-based results can thus be extended to non-lens galaxies {\sl in the same parameter space} \citep[e.g.][]{2008arXiv0810.4924H, 2008arXiv0808.2497M}.

In this letter we study the total mass-density profile of massive early-type galaxies inside one effective radius, using the {\sl full} SLACS sample of 58 gravitational lens systems with high-fidelity HST-ACS observations. We examine the intrinsic scatter in density slopes that is allowed by the sample, whether the slope correlates with other global parameters and we set limits on the level of orbital anisotropy in these system. 
In Section 2, we present the logarithmic density slopes of the SLACS early-type lens galaxies, based on two different methods, one based on lensing and dynamics and one based on scaling relations, that explicitly include their average density profile as a free parameter. Comparing these two values allows us to set limits on their average orbital anisotropy. In section 3, we summarize our results and conclusions. Throughout this paper, we make use of the sample of 58 SLACS single-lens systems from \citet{2008ApJ...682..964B} and take all quantities from that paper.  If not mentioned otherwise, all masses are in units of $10^{10}$M$_{\odot}$ .  We assume $\Omega_{\rm m}=0.3$, $\Omega_{\Lambda}=0.7$ and $H_0=100\, h$~km/s/Mpc.

\section{The Density Profile of Massive ETGS}

To determine the logarithmic slope of the total density profile, we use two alternative methods: (i) through combining SDSS-based stellar velocity dispersions and lensing-based total masses, and (ii) through scaling relations between luminosity, mass and rescaled radius, which does {\sl not} require a measured stellar velocity dispersion. Whereas a single-power law mass model is an approximation, \citet{2003ApJ...583..606K, 2004ApJ...611..739T, 2009MNRAS.393.1114B} show that two component models can robustly be approximated by a single power-law component given current data quality.   

\subsection{Derivation from Lensing and Stellar Dynamics}

To derive the logarithmic density slopes, we follow \citet{2002ApJ...575...87T, 2003ApJ...583..606K,  2004ApJ...611..739T, 2006ApJ...649..599K}.  First the lensing mass inside the critical curves is determined from the lens models in \citet{2008ApJ...682..964B}, which is nearly invariant under changes in the density profile \citep{1991ApJ...373..354K}, hence the assumed density profile during this step is not relevant in further steps (see \citet{2006ApJ...649..599K}). Subsequently we solve the spherical Jeans equations -- assuming this mass as external constraint and an Einstein radius equal to the deflection angle of an equivalent spherical mass distribution -- for a luminosity-density profile that follows either a \citet{1990ApJ...356..359H} or \citet{1983MNRAS.202..995J} profile, embedded as trace-component inside the total density profile $\rho_{\rm tot} \propto r^{-\gamma'_{\rm LD}}$ (i.e. a power-law profile). The half-light radius of the projected luminosity-density profile is set equal to the observed effective radius \citep{2008ApJ...682..964B}. We take seeing (FWHM=1.5 arcsec) into account.  We vary the slope $\gamma'_{\rm LD}$ over a range of 1.1 to 2.9 and compare the predicted velocity dispersion inside the 3-arcsec diameter SDSS fiber with the observed value. The error in the measured velocity dispersion is by far the most dominant source of uncertainty.  Hence, the likelihood $dP/d\gamma'_{\rm LD} \propto e^{-\chi^2/2}$ is determined from the $\chi^2$ mismatch between the model and observed velocity-dispersion values.

\begin{figure}
\begin{center}
\resizebox{\hsize}{!}{\includegraphics{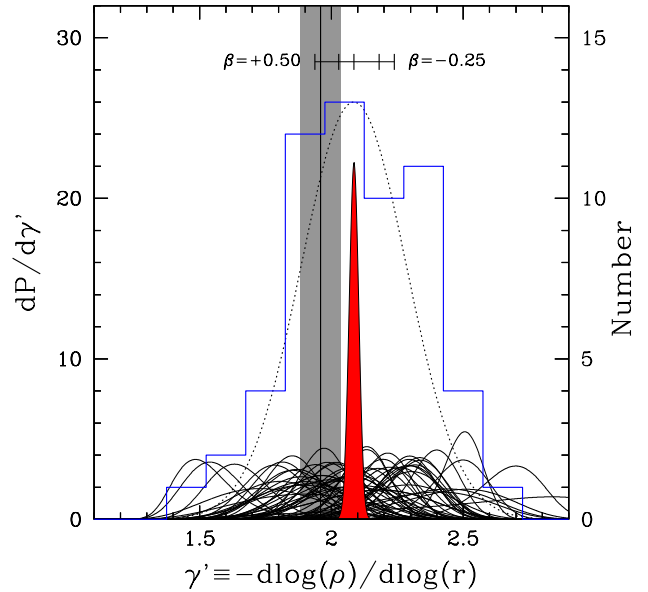}}
\caption{The logarithmic density slopes of 58 SLACS early-type galaxies (thin solid curves).
The filled red curve is the joint likelihood of the {\sl ensemble-average} density slope. The histogram indicates the distribution of median values and the dotted Gaussian curve indicates the intrinsic scatter in $\gamma'_{\rm LD}$ (see text for details). We assume a Hernquist luminosity-density profile. 
The small dashes indicate the shift in  the ensemble-average density slope 
for $\beta_r=+0.50, +0.25, -0.50 -0.25$ (left to right),
respectively. Note the reversal of the $\beta_r=-0.50$ and $-0.25$ dashes. The vertical solid line and gray region
indicates the best-fit value and 68\% CL interval, respectively, of the average density derived from scaling relations.
 }
\label{fig:slopes}
\end{center}
\end{figure}

The results in Figure~\ref{fig:slopes}  show that most values lie around a slope of two, which is that of an isothermal mass profile ($\rho \propto r^{-2}$).
A joint analysis of the sample yields:
\begin{equation}
	\langle \gamma_{{\rm LD}}' \rangle \equiv -d\log(\rho_{\rm tot})/d\log(r)=2.085^{+0.025}_{-0.018} {\rm ~~(68\% ~CL)},
\end{equation}
in the range of radii $0.2 - 1.3$\,$R_{\rm eff}$ and for $\beta_r=0$. The dependence on orbital anisotropy is small with the slopes varying mildly over $\beta_r=\pm 0.50$ (see Fig.\,\ref{fig:slopes}).  Based on changing the luminosity density profile, seeing, etc., we estimate an $\sim$0.1 systematic error. We note two points: (i) a more detailed two-dimensional kinematic analyses of six SLACS galaxies \citep[see][]{2008MNRAS.384..987C, 2009arXiv0904.3861B} agree with these results, and (ii) comparing the density slopes of the 14 systems that overlap with \citet{2006ApJ...649..599K}, we find an average increase of $\gamma'_{\rm LD}$ by 6\%. This difference can be attributed to minor model improvements, the use of better HST images, leading
to an average decrease of $R_{\rm eff}$ by 13\%, and
an improved derivation of the stellar velocity dispersion, leading to an increase by +3\%. In particular the latter leads to an average increase in $\gamma'_{\rm LD}$, explaining most of the difference.
Hence currently we are limited by systematics.

An intrinsic spread of $\sigma_{\gamma'}=0.20^{+0.04}_{-0.02}$ (i.e.\ $\sigma_{\gamma'}/\langle \gamma'_{\rm LD} \rangle = 0.10^{+0.02}_{-0.01}$; 68\% CL) is derived, assuming Gaussian intrinsic and error distributions \citep[see][]{2006ApJ...649..599K, 2009arXiv0904.3861B} consistent with the scatter found in \citet{2006ApJ...649..599K, 2007ApJ...671.1568J, 2009arXiv0904.3861B}. Despite the uniformity of the sample, {\sl differences between galaxies are present}, which could partly be physical \citep[see e.g.][]{2001AJ....121.1936G}, partly due to systematics, or due to small uncorrelated contributions from the environment and large scale structure \citep[][]{2008MNRAS.383L..40A, 2009ApJ...690..670T, 2009arXiv0904.4381G}. Conservatively it should therefore be regarded as an upper limit on {\sl physical} variations.

\subsection{Derivation from Scaling Relations}

A second method to derive the ensemble-average density profile is to assume a scaling relation between the observables, luminosity, effective radius and Einstein radius and Einstein mass:
\begin{equation}
	\alpha \log(L_{\rm eff}) = \log\left[M_{\rm einst}
\left(\frac{R_{\rm eff}}{R_{\rm einst}}\right)^{(3-\gamma'_{\rm SR})}\right] + \delta 
\end{equation}	 
This relation assumes that the density profile scales as a power-law with $\rho\propto r^{{-\gamma'_{\rm SR}}}$ and that mass and light scale with $M_{{\rm eff}} \propto L_{{\rm eff}}^{\alpha}$ with $M_{{\rm eff}} \equiv M_{\rm einst} \left({R_{\rm eff}}/{R_{\rm einst}}\right)^{(3-\gamma'_{\rm SR})}$. The idea is that for a fixed luminosity, $M_{\rm einst}$  scales with $R_{\rm einst}/R_{\rm eff}$ and thus provides an ensemble-averaged density slope.

This approach is slightly different from \citet{2008ApJ...682..964B}, who assumed a SIE or constant M/L mass profile
in deriving the MFP, but similar to \citet{2005ApJ...623..666R} who focused on the FP and density slope. It allows us to derive the density slope {\sl independently} from assumptions about stellar dynamics. The resulting values are $\alpha = 1.363 \pm 0.056$, $\delta=-0.959 \pm 0.050$ and for the best-fit ensemble-average logarithmic density slope
\begin{equation}
	\langle \gamma'_{{\rm SR}} \rangle = 1.959 \pm 0.077.
\end{equation}
This result is close to that derived based on dynamics models, although it assumes nothing about the dynamical structure of these galaxies (e.g. isotropy). The value of $\alpha\equiv 1/\eta'$, with $\eta'$ as defined and given in \citet{2008ApJ...682..964B}. The difference between $\langle \gamma'_{{\rm SR}}\rangle$ and $\langle \gamma'_{{\rm LD}}\rangle$ of $\sim 0.1$ implies that on average anisotropy can not be very large (see \S\ref{sect:anisotropy}).

\subsection{Correlations of Slope and Galaxy Properties}

To assess whether $\gamma'_{\rm LD}$ correlates with global galaxy quantities or cosmic time, we plot them against effective mass, rescaled ($R_{\rm einst}/R_{\rm eff}$) radius and redshift.  The results are shown in Figure~2.  We find the following linear gradients:
\begin{equation} 
\left\{ 
\begin{array}{l}
{d \gamma'_{\rm LD}}/{d \sigma_{\rm SIE}}  =  (-1.6 \pm 8.3) \cdot 10^{-4} \\ 
~\\
{d \gamma'_{\rm LD}}/{d M_{\rm eff}} = (-0.6 \pm 14.5)\cdot 10^{-4} \\
~\\
{d \gamma'_{\rm LD}}/{d (R_{\rm einst}/R_{\rm eff})} =  (0.10 \pm 0.16)\\
~\\
{d \gamma'_{\rm LD}}/{d z} = (-0.05 \pm 0.43),
\end{array}
\right. 
\end{equation}
with $\sigma_{\rm SIE}$ in units of km/s and $M_{\rm eff}$ in units of $10^{10}$\,M$_{\odot}$. No weights on
the points are used to avoid the brighter low-redshift galaxies from dominating the fits. We find no correlations at any significant level. Similar results were found by \citet{2006ApJ...649..599K} and recently by \citet{2008MNRAS.383L..40A, 2009ApJ...690..670T, 2009arXiv0904.4381G}, based on studies of their environment
and mass along the line-of-sight. Overall, the inner regions ($R\la R_{\rm eff}$) of massive early-type ($>L_*$) galaxies are remarkably homologous and simple. This result is unlike some of the more detailed results found by the SAURON collaboration for $\la L_*$ galaxies \citep[e.g.][]{2006MNRAS.366.1126C,  2007MNRAS.379..401E}, which show a range of different kinematic signatures (e.g.\ counter-rotating cores, triaxiality, fast versus slow rotators, etc.). We find that these complexities, at least at the high-mass end, does not seem to affect the derivation of their mass distributions and scaling relations \citep[see e.g.][for a discussion]{2009MNRAS.393.1114B}, although the overlap between the samples in mass is small. The analyses of six SLACS lens system \citep{2008MNRAS.384..987C, 2009arXiv0904.3861B}, based on VLT VIMOS-IFU data, shows that the majority of the SLACS galaxies are slow rotators predominantly supported by random stellar motions.

\begin{figure}
\begin{center}
\resizebox{\hsize}{!}{\includegraphics{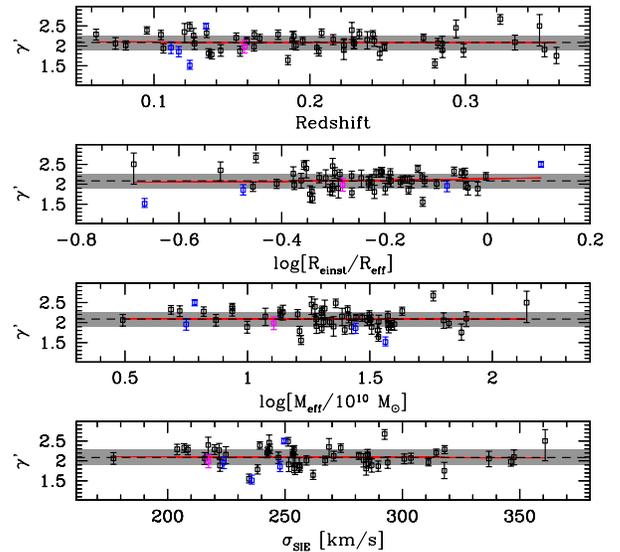}}
\caption{The median values of $\gamma'_{\rm LD}$ versus global galaxy quantities and redshift.
The blue symbols are S0 galaxies and the magenta colored system is an E/S0 galaxy. The dashed curves is given 
at $\gamma'_{\rm LD} \equiv 2.085$ for reference with the gray box being $\pm$10\% intrinsic
scatter. The thin red line is the best linear fit (i.e. shown curved in the log-plots).}
\label{fig:slopecorrelations}
\end{center}
\end{figure}

\subsection{Limits on Orbital Anisotropy}\label{sect:anisotropy}

Of the two methods to derive the average density slope, one
depends on dynamics and the other does not. In the former, we
assumed $\beta_r=0$. If the galaxies, however, on average have a different value of $\beta_r$,
the average value of $\gamma'_{\rm LD}$ can be either too large or to small, since we would attribute a
higher/lower dispersion to a steeper/shallower density slope and not to radial/tangential orbital anisotropy.

The independence of the second method from dynamics allows us therefore to place constraints on the average 
orbital anisotropy of the ensemble of galaxies $\langle \beta_r \rangle$. As indicated in Fig.1, good agreement is
found for a positive value of $\langle \beta_r \rangle \approx 0.45 \pm 0.25$,
in agreement with results from \citet[][]{2001AJ....121.1936G}. The galaxies are at most mildly radial anisotropic, as also found in \citet[][]{2008MNRAS.384..987C, 2009arXiv0904.3861B} from a more detailed two-integral and two-dimensional analysis of SLACS systems.

\section{Results \& Conclusions} 

We have presented a complete analysis of the inner mass density profile of 58 early-type galaxies from the SLACS survey, focusing on their logarithmic density slope. We find the follow results for galaxies with $M_{\rm eff} \ga 3\cdot 10^{10}$\.M$_{\odot}$:

\begin{itemize}
        \item Based on lensing and stellar dynamics constraints, the {\sl average} inner logarithmic density slope in their inner 0.2--1.3 R$_{\rm eff}$
         is $\langle \gamma'_{\rm LD} \rangle = 2.085^{+0.025}_{-0.018}$ (random error on mean; 68\% CL; for isotropic orbits, $\beta_r=0$),
	$\pm 0.1$ (syst.) and a small ($\la$0.1) dependence on anisotropy for a range of $\beta_r= \pm 0.50$. 
	   An intrinsic scatter in $\gamma'_{\rm LD}$, between
	individual galaxies, of $\sigma_{\gamma'}\la 0.20^{+0.04}_{-0.02}$ (68\% CL)
	is still allowed by our data. This should be regarded as an
	upper limit on physical variations, since it must include residual systematic effects. However, we believe part of it
	   to be due to real differences between ETG density profile slopes \citep[see also][]{2009MNRAS.393.1114B}.
	
	\item A second independently-derived value of the {\sl average} inner logarithmic density slope
	is  found from scaling relations, yielding $\langle \gamma' _{\rm SR}\rangle =1.959\pm 0.077$.
	This value is completely independent from dynamics and therefore a robust sanity
	check of the lensing plus dynamics results.
	
	 \item Since  $\langle \gamma' _{\rm SR}\rangle$ is independent of orbital anisotropy, we 
	  can set a weak limit $\langle \beta_r \rangle = 0.45 \pm 0.25$ on the average anisotropy of these systems.      
	    This shows  that massive galaxies are at most mildly radial anisotropic.
	
	\item No correlation of $\gamma'_{\rm LD}$ is found with either galaxy mass or redshift,
	nor with radius over which this slope is measured (0.2--1.3 R$_{\rm eff}$), 
	implying that these results are genuine and widely applicable.

	\item Taking the density slope into account, an increase in luminosity is found with increasing
	galaxy mass: $M_{\rm eff}\propto L_{\rm eff}^{1.363 \pm 0.056}$ in agreement with \citet[][]{2008ApJ...684..248B}.
	
\end{itemize}

Based on these numerical results, we conclude that massive early-type galaxies with total masses $M_{\rm eff}\ga 3 \cdot 10^{10}$M$_{\odot}$ are structurally and dynamically very similar in their inner regions (roughly one effective radius), over a look-back time of about 4\,Gyr,  with stellar and dark-matter adding up to a combined close-to isothermal density profile but having some, although relatively little, intrinsic scatter between their logarithmic density slopes. This {\sl bulge-halo conspiracy} occurs despite (i) a clear increase in their total $M/L$ inside one effective radius with galaxy mass \citep[as found in this paper and in][]{2008ApJ...684..248B}, leading to a tilt in the FP, and (ii) a very complex hierarchical formation history. Moreover, from the agreement between two different determinations of their average density slope -- one dependent on dynamics and the other not --  we also find that these galaxies are also at most mildly radially anisotropy. Our results are based on single-component mass models, which provide  
a good description of the available data for systems. In forthcoming publications, we will address two-component models
where stellar and dark matter are modeled separately.

Our results on the near homology, isothermality and isotropy of massive early-type galaxies provide new  constraints on theoretical models and numerical simulations of the formation of early-type galaxies, their subsequent evolution and the understanding of the FP. Whereas these models/simulations should match these scaling relations, they should also
match their intrinsic scatter. 

\acknowledgements

L.K. is supported through an NWO-VIDI program subsidy. T.T. acknowledges support
from the NSF through CAREER award, by the Sloan and Packard Foundations. The work of LAM was carried out at JPL/Caltech, under a contract with NASA. Support for HST programs \#10174, \#10587, \#10886, \#10494, \#10798, \#11202 was provided by NASA through a grant from the STScI.

\bibliographystyle{apj}

\end{document}